\documentclass[12pt]{article}

\usepackage{latexsym}

\begin{document}


\title{ Solar-System Experiments and Saa's Model of Gravity
with Propagating Torsion}

\author{
        Fiziev P.\thanks{E-mail: fiziev@phys.uni-sofia.bg}\,\,,
     Yazadjiev S.\thanks{E-mail: yazad@phys.uni-sofia.bg}\\
\\
{\footnotesize  Department of Theoretical Physics,
Faculty of Physics,}\\
{\footnotesize Sofia University,}\\
{\footnotesize 5 James Bourchier Boulevard, Sofia~1164, }\\
{\footnotesize Bulgaria }\\
}

\maketitle

\begin{abstract}
It's shown that  Saa's model of gravity with propagating torsion is
inconsistent with basic solar-system gravitational experiments.
\end{abstract}


\sloppy
\renewcommand{\baselinestretch}{1.3} %
\newcommand{\sla}[1]{{\hspace{1pt}/\!\!\!\hspace{-.5pt}#1\,\,\,}\!\!}
\newcommand{\db}{\,\,{\bar {}\!\!d}\!\,\hspace{0.5pt}}
\newcommand{\partb}{\,\,{\bar {}\!\!\!\partial}\!\,\hspace{0.5pt}}
\newcommand{\dsla}{\partb}
\newcommand{\eql}{e _{q \leftarrow x}}
\newcommand{\eqr}{e _{q \rightarrow x}}
\newcommand{\ite}{\int^{t}_{t_1}}
\newcommand{\itz}{\int^{t_2}_{t_1}}
\newcommand{\itd}{\int^{t_2}_{t}}
\newcommand{\lfrac}[2]{{#1}/{#2}}
\newcommand{\sfrac}[2]{{\hbox{${\frac {#1} {#2}}$}}}
\newcommand{\dV}{d^4V\!\!ol}
\newcommand{\ben}{\begin{eqnarray}}
\newcommand{\een}{\end{eqnarray}}
\newcommand{\la}{\label}


Recently a new model of gravity involving propagating torsion was proposed
by A. Saa \cite{Saa1}-\cite{Saa5}.
In the present note we investigate the consistency of this model
with the basic solar-system gravitational experiments.

The main idea of the model is to replace the usual volume element
$\sqrt{-g}d^4x$ with a new one  --  $e^{-3\Theta}\sqrt{-g}d^4x$,
which is covariantly constant with respect to the transposed affine connection
$\nabla^{T}$.
By virtue of the using a new volume element the equations of the motion
for the matter and gauge fields are of an autoparallel type
\cite{Saa1}-\cite{Saa5}, \cite{F}.

In presence of  spinless matter only in this model an Einstein-Cartan
geometry with semi-symmetric torsion tensor
$S_{\alpha\beta}{}^{\gamma}=S_{[\alpha}\delta_{\beta]}^{\gamma}$
is considered, where the torsion vector $S_{\alpha}$ is potential,
i.e. there exists a potential $\Theta$ such that $S_{\alpha}=
\partial_{\alpha}\Theta$.
The following new equations for geometrical fields (metric and torsion)
are obtained \cite{Saa1}-\cite{Saa5}, \cite{F}:
\ben
G_{\mu\nu} +\nabla_\mu \nabla_\nu \Theta - g_{\mu\nu}\Box\Theta =
{\sfrac \kappa {c^2}}T_{\mu\nu},\nonumber\\
\nabla_\sigma S^\sigma= \Box\Theta= -{\sfrac {2\kappa} {c^2}}
\left({\cal L}_M -{\sfrac 1 3}{\delta{\cal L}_M \over \delta \Theta}
+{\sfrac 1 2}T\right).
\la{SYS}
\een
Here $G_{\mu\nu}$ is the Einstein tensor for the affine connection $\nabla$,
${\cal L}_M$ is the matter lagrangian density,
$T$ is the trace of the energy-momentum tensor $T_{\mu\nu}$,
and $\kappa$ is the Einstein gravitational constant.

It is not hard to see that the equations (\ref{SYS}) for geometric fields
$g_{\alpha\beta}$ and $\Theta$ in vacuum coincide with the corresponding
equations in Brans-Dicke theory \cite{BD}, \cite{Brans} in vacuum with
parameter \,$\omega= -{\sfrac 4 3}$\, if the field $\Theta$ in the Saa's model
is replaced with a Brans-Dicke scalar field $\Phi = e^{-3\Theta}$.
Hence,
the asymptotically flat, static and spherically symmetric general solution
of the vacuum geometric fields equations (\ref{SYS}) is known \cite{Brans},
\cite{ZX}.
In isotropic coordinates it's given by a
two-parameter -- $(r_{0},k)$  family of solutions
\ben
ds^2 =
\left(1-{r_{0}\over r}\over 1+{r_{0}\over r}\right)^{2\over \rho(k)}(c\,dt)^2
- \left(1 - {r_{0}^2 \over r^2}\right)^2
\left(1-{r_{0}\over r} \over 1+{r_{0}\over r}\right)^{{2\over \rho(k)}(3k - 1)}
\left(dr^2  + r^2d\Omega^2\right) ,
\la{metric}
\een
where $\rho(k)=\sqrt{3\left(k - {1\over 2}\right)^2 + {1\over 4}}$.

The parameter $k$ presents the ratio of the torsion force
(as defined in \cite{F}) and the gravitational one \cite{BFY}.
In the case when $k = 0$ we have the usual torsionless Schwarzschild's
solution and $r_0 \equiv 4 r_g$ is the standard gravitational radius.

The week field approximation used in \cite{BD} yields the formal value
$k = {1 \over 2}$ when the source of the geometrical fields (metric and torsion)
is a point with mass $M$, but this approximation
is not valid in general for the case of a star \cite{BD}.
Moreover, in Saa's model no result like Birkhoff theorem
in general relativity take place.
Therefore it is impossible to use all results for fields in vacuum
obtained in the case of a point source for the fields in vacuum
when the source of these fields is a massive star. Hence, we are not able
to reject immediately Saa's model using the well known facts
about the relation of the Branse-Dicke theory with solar-system experiments.
A further study of the problem is needed.

The asymptotic expansion of the metric (\ref{metric}) at
$r\rightarrow \infty$ is
\ben
ds^2 \approx \left(1 - {4r_{0}\over \rho(k)r} +
{8r_{0}^2\over \rho(k)^2 r^2}\right)(c\,dt)^2
- \left(1 + (1 - 3k){4r_{0}\over \rho(k)r}\right)
\left(dr^2  + r^2d\Omega^2\right) .
\la{As1}
\een
In the model under consideration, as it was shown in \cite{F},
the test particles move along  geodesic lines.
Therefore the active gravitational mass
"seen" by the test particles is $$M={2r_{0}\over \rho(k)}$$ and
correspondingly the asymptotic expansion (\ref{As1})
may be written in the form
\ben
ds^2 \approx \left(1 - {2M\over r} + 2{M^2 \over r^2} \right)(c\,dt)^2
- \left(1 + 2(1 - 3k){M\over r}\right)\left(dr^2  + r^2d\Omega^2\right).
\la{As2}
\een
From here, it immediately follows that the first two of the post-Newtonian
parameters are
$$\beta = 1, \gamma = 1 - 3k.$$

The solar-system experiments set tight constrains on the post-Newtonian
parameters \cite{Will}:
$$\mid\beta - 1\mid < 1*10^{-3} , \mid\gamma - 1\mid < 2*10^{-3}.$$
Therefore  Saa's model will not contradict to the experimental facts
if the parameter $k$ satisfies the inequality:
$$\mid k\mid < {2\over 3}*10^{-3}.$$

In order to specify the theoretically possible values of $k$ in Saa's model
we must consider a model of a star with mass $M$ as a source of
geometrical fields\footnote{In presence of matter
the field equations of this model differ essentially from the corresponding
equations in Branse-Dicke theory. Hence, an independent consideration of
a star in Saa's model is needed}.
Such a model was investigated in \cite{BFY}.
It was shown that the parameter $k$ satisfies the constraint
$${1\over 3}\leq k \leq {1\over 2}$$
under general physical assumptions: 1) spherically symmetric and stationary
star state; 2) regularity of the solution at the center of the star; and
3) validity of the positivity condition $T = \varepsilon -3p \ge 0$,
$\varepsilon$ being the  energy density of the star, $p$ being the pressure.
The last condition is critical and means that the star is build of a normal
matter. The parameter $k$ takes maximal value ${1 \over 2}$
(which coincide with its value obtained in the week field limit)
in the case of nonrelativistic matter $\varepsilon \gg p$
and minimal value ${1 \over 3}$ in the ultrarelativistic case
$\varepsilon = 3p$.

Consequently  Saa's model is inconsistent with basic solar-system
gravitational experiments\footnote{For a matter with unusual properties when
$\varepsilon \approx p $ (i.e. $T = \varepsilon -3p < 0$ \cite{Zel1}, \cite{Zel2})
we have $k \approx 0$ and Saa's model may not contradict to observations,
but this is not the case of the solar system.}.

\bigskip

\noindent{\Large\bf Conclusions}
\bigskip

Saa's model seems to be interesting because
in the metric-affine-connected spaces with nonzero torsion and curvature
it leads to a consistent application of the minimal coupling principle
both in the action principle and in the equations of motion for fields
(the last being of an autoparallel type) \cite{Saa1}-\cite{Saa5}.
This property of the model is not valid in the same spaces when one
applies the minimal coupling principle both in the action
principle and in the equations of motion (of a geodesic type)
for particles \cite{F}.
Most probably the last inconsistency is responsible for the negative result
we obtained in the present note.
Thus we see that to comply  Saa's model with the experimental data,
if this is possible at all, we must try to change properly this model.
\bigskip

\noindent{\Large\bf Acknowledgments}
\bigskip

This work has been partially supported by
the Sofia University Foundation for Scientific Researches, Contract~No.~245/98,
and by
the Bulgarian National Foundation for Scientific Researches, Contract~F610/98.

\end{document}